\documentclass[runningheads]{llncs}
\usepackage[T1]{fontenc}
\usepackage{graphicx}

\begin{document}

\title{An XAI Approach to Deep Learning Models in the Detection of DCIS}
\author{Michele La Ferla\inst{1}\orcidID{0000-0002-2230-6305}\\
Matthew Montebello\inst{1}\orcidID{0000-0002-6177-7548}\\
Dylan Seychell\inst{1}\orcidID{0000-0002-2377-9833}}
\authorrunning{M. La Ferla}
\institute{University of Malta, Msida MSD 2080, Malta\\
\url{https://www.um.edu.mt/ict/ai}
\email{michele.laferla.05@um.edu.mt}}

\maketitle

\begin{abstract}
Deep Learning models have been employed over the past decade to improve the detection of conditions relative to the human body and in relation to breast cancer particularly. However, their application to the clinical domain has been limited even though they improved the detection of breast cancer in women at an early stage. Our contribution attempts to interpret the early detection of breast cancer while enhancing clinicians’ confidence in such techniques through the use of eXplainable AI.

We researched the best way to back-propagate a selected CNN model, previously developed in 2017; and adapted in 2019. Our methodology proved that it is possible to uncover the intricacies involved within a model; at neuron level, in converging towards the classification of a mammogram. After conducting a number of tests using five back-propagation methods, we noted that the Deep Taylor Decomposition and the LRP-Epsilon techniques produced the best results. These were obtained on a subset of 20 mammograms chosen at random from the \textit{CBIS-DDSM} dataset. The results showed that XAI can indeed be used as a proof of concept to begin discussions on the implementation of assistive AI systems within the clinical community.

\end{abstract}


\section{Introduction}
Breast carcinoma is the most common types of tumours among women in the western world, according to WHO reports \cite{Who2019}. As a result, it is very common that many studies have devoted significant time and effort in developing models that can aid radiologists obtain accurate and timely diagnoses of this disease. The mammographic presence of breast cancer can initially be noted through one of the following four  methods: by discovering minor distortions of the breast tissue; showing the presence of masses in the breast; through the presence of non-asymmetrical breasts; or through the presence of microcalcifications. 

We started our study by collecting information from a questionnaire which was distiributed to 12 of the 15 clinical specialists in breast cancer, which Malta is currently equipped with. This helped shed light about their practices, as well as their attitudes towards the employment and effectiveness of AI technologies. The similarity of our findings to international literature encouraged us further to pursue our focus on addressing the lack of implementation of assistive AI models in hospitals and also investigate how the introduction of XAI could help clinicians understand how a model classifies between a benign, malignant, or non-tumorous breast tissue.

This was followed by a comparative study between different available datasets to adopt as part of our scientific tests that employed a Convolutional Neural Network (CNN). The \textit{CBIS-DDSM} scanned film mammographic dataset was selected due to its unique features, being that is has been carefully annotated by expert radiologists and has also been extensively used in the deep learning community. Additionally, we used this dataset on an already trained model, due to a lack of computational resources, and to further focus our study on the element of XAI in deep learning.

The results obtained from reverse engineering a CNN model using back-propagation methodologies are discussed further on in this paper and were also presented to the same clinicians interviewed in the initial questionnaire. Optimal results were obtained when using the Deep Taylor Decomposition and LRP-Epsilon techniques. The best performing solution would be a combination of these two techniques to achieve ideal results.

\section{Literature Review}
In the early 1990s, one of the first datasets on breast cancer was published on the Machine Learning Repository site. This contained information from a total of 699 patients diagnosed with breast cancer. Although very primitive when compared to the mammogram screening image datasets we have today, the dataset has been extensively used in several projects in the following years by utilising its features to predict whether people with similar identifiable traits are more likely to be diagnosed with a tumour. The dataset was curated by the University of Wisconsin hospital in the US and contained data for women who were investigated for breast cancer covering the period between 1989 and 1992 \cite{10.5555/221982}.

\subsection{Research Studies Involving Deep Learning}
Maybe one of the greatest ground-breaking studies made in this research area was the one made by the Google Deepmind team, in collaboration with Cornell University. The research made here involved the creation of the \textit{LYNA} model which helped in the diagnosis of breast cancer through MRI images, which included large pathology images for lymph nodes. The purpose of This study was to identify the presence of a tumour in the lymph nodes around the breast area of female patients. The dataset used in this study involved the analysis of images which have portrayed biopsies from MRI stage procedures in the second stage of a diagnosis and used to develop a deep learning model able to detect a tumour which is at a metastasis stage \cite{stumpe}. While the LYNA model proved to be significantly more effective in detecting breast cancer, the researchers themselves admitted in their study that an accurate model alone does not prove enough to improve the diagnostic work done by pathologists or improve outcomes for breast cancer patients \cite{liu2017detecting}. A very important outcome of this study was that of keeping in mind patient safety using machine learning techniques on patients. Such a model would need to be tested in different scenarios, and use diverse datasets to understand its predictive power. Furthermore, the benefits of such a system whereby medical practitioners used the \textit{LYNA} model had still to be explored. It was still too early to determine whether such an algorithm improved the efficiency of the procedure or diagnostic accuracy. The importance of \textit{LYNA} was however proved in two other studies carried out in 2018 and 2019, respectively \cite{khan2019AND,steiner}.

During the NIPS17 conference, one particular study made by Shen et al. contributed significantly to tumour identification using screening mammogram images \cite{li_2019}. This was later improved upon in 2019 by the same researchers \cite{Rothstein}. The study initially based its model on the Yaroslav architecture in 2017, but was then tweaked to compare the VGG-16 and ResNet-50 architectures, in an attempt to improve the detection of breast tumours by reading mammograms. The second model modified the first to use a fully convolutional training approach which effectively makes use of the annotations on the curated \textit{CBIS-DDSM} dataset. By using this approach, certain dataset features such as lesion annotations are only needed for the initial training. The inner layers within the model would only require image-level labels, therefore decreasing the reliance on lesion annotations. This was a significant improvement in the radiological field of study, since the availability of annotated mammogram datasets is even to date very limited \cite{Rothstein}. The relevance of this project is that it is part of an open-source study which can therefore be used and altered to improve on by other researchers \cite{li_2019}. Shen et al. used the publicly available well-curated \textit{CBIS-DDSM} and inBreast mammographic datasets of scanned film mammography studies to build their model upon \cite{usfdigitalmammographyhomepage}.

\subsection{Developments in XAI}

The best approach to consider was that of using Layer-wise Relevance Propagation, which in theory works backwards through the neural network to redistribute the output result back to all the neurons (or pixels) in the input image. The redistribution process can be simplified using formula 1, which uses back propagation from one layer to the previous:

\begin{equation}
	R_j = \sum_k \frac{x_j w_{j,k}}{\sum_j x_j w_{j,k} + \epsilon} R_k
\end{equation}

The function of the formula above is to use neuron activators and weight connections to interpret a deep learning model. This is particularly relevant to ResNet-50 architectures due to their residual properties. \(x_j\) takes the value of the activator for the neuron \(j\) in any layer of the network. \(w_{j,k}\) is used as a weighting given to the connection between neuron \(j\) and neuron \(k\) in the next one. \(R_j\) has the property of the relevance score for each neuron in the first given layer, and \(R_k\) is the relevance score for each neuron in the next inner layer \cite{Binder}.

LRP is considered a rather conservative algorithm to backpropagate a deep learning model. This means that in essence the magnitude of any output \(y\) is aligned throughout the back-propagation process and is equivalent to the sum of the relevance map \(R\) in the input layer. This property is true for any hidden consecutive layers \(j\) and \(k\) within the neural network and transitivity for the input and output layer \cite{lindwurm_2019}.

The numerator of the fraction in the formula is the value at which the neuron \(j\) can influence the neuron \(k\), which is valid for the linear case of an active ReLU activator. This is split up by the aggregate of contributions in all lower-layer neurons for us to enforce the conservation property. The outer sum over \(k\) is a representative of the relevance of the neuron \(j\). It is calculated using the sum of its influence on all neurons \(k\) from the following layer and multiplied by the \(R\) value of these neurons.

\section{Materials \& Methods}

After developing the initial patch classifier for the NIPS-16 competition Shen et al. moved on to improve on their previous model and built a whole image classifier rather than the previous patch-classifier. The advantage of this is that the new model didn't segment the image into patches but processed the mammogram as a whole. To do this they flattened the heatmap and connected it to the image classification output using a novel idea called fully connected layers. A max-pooling layer followed the fully connected layer and was used to partially eliminate the imbalance brought by the translational invariance which the previous version of the model suffered with \cite{li_2019}. In addition, a shortcut was also introduced between the heatmap and the output layer to facilitate training. The equation used to achieve this result using softmax activation is shown in formula 2. 

\begin{equation}
	f(z)_j = \frac{e^{z_j}}{\sum^{c}_{t=1}{e^{z_t}}} \mathrm{for} j = 1, \ldots, c
\end{equation}

\subsection{Layer-Wise Relevance Propagation}

Following the research by Montavon and Binder \cite{Montavon,Binder}, it was found that the Layer-wise Relevance Propagation method is one of the most effective algorithms used to explain and interpret decisions in deep learning networks. Within the ResNet-50 CNN architecture, the explanation given by LRP can be represented through back-propagation utilising those pixels contained in the image in question which influence the outcome as to how the model classifies the image.

The primary benefit of this technique is that it does not conflict with network training, so it could be independently applied to the already trained classifiers in any dataset. Based on a second study by Lehman et al., it is hoped that LRP could provide physicians with the tools not only to interpret mammograms but to alert them to the presence of other co-diseases such as tumours and possible cardiovascular disease \cite{Regina}. Researchers and medical professionals alike are always enthusiastic to try new methods, particularly those with a lower risk to the patient.

\subsection{Deep Taylor Decomposition}

Over and above the work done by Montavon and Binder on LRP \cite{Montavonbook} at the same time, the same authors coined the Deep Taylor decomposition method to better analyse the interpretability of deep learning models. This is an improvement on the original LRP methodology since the authors found a number of constraints, from which one could derive different functions; one of which being Deep Taylor decomposition. If we were to decompose the function \(f\) in the equation below in terms of its partial derivatives, the result can be used to approximate the relevance propagation function. The closer \(x\) is to \(x_0\), then the better the approximation. This simplifies the Deep Taylor equation to the following one \cite{shiebler}.

\begin{equation}
	f (x) \approx {\sum_{d=1}^{v}} \frac{\partial f}{\partial x_{(d)}}(x_0)(x_{(d)} - x_{0(d)}) \mid f (x_0) = 0
\end{equation}

The Deep Taylor decomposition method helps us understand which neurons which contribute to the classification of an image and work best with mono-chromatic images such as the ones used in mammograms, X-rays and other medical images. 

\subsection{Comparing the Regions of Interest}

The five chosen techniques used for comparison, all formed part of the iNNvestigate library and are the following:

\begin{itemize}
	\item Deconvolution Network \cite{Alber}
	\item Deep Taylor Decomposition \cite{Montavonbook}
	\item LRP-Preset A Flat \cite{Alber}
	\item LRP-Epsilon \cite{Alber}
	\item Guided Backprop \cite{Reyes}
\end{itemize}

The decision to use these methodologies, in particular, was based on a number of previous studies whereby they were used to interpret medical images; some of which mammograms in particular. Notable among these is the research made by Reyes et al. in 2020 about the interpretability of medical images in radiology. In their attempt to discuss several backpropagation models, the authors compare the Guided Backprop method to others such as gradient-weighted class activation mapping, pointing out that the Guided Backprop methodology performs well when interpreting MRI images of the human brain \cite{Reyes}.

\section{Results and Discussion}

When comparing the five different methodologies we tested to determine the one which interpreted mammograms best, we noticed that the Deep Taylor Decomposition did in most cases perform better than the other four, however, there were instances where the LRP-Epsilon technique gave better results. In the following section, we will go through the different results achieved on the sample subset of 20 mammograms from the \textit{CBIS-DDSM} dataset.

The first notation we may take from the subset used to test the five different methods is that the Guided Backprop and the LRP Preset A-Flat methods are not helpful for our purposes. Starting with the Guided Backprop method, we noticed that this tends to scatter the pixels all over the image, losing the important parts of the mammogram which identify its classification. There were some instances in the subset, where the Guided Backprop method gave certain information about regions in the mammogram where abnormality may occur; such as in 016, 223 and 242; where there is a more dense amount of pixels around particular regions of the mammograms. However, having said that, we found that other methods performed better when attempting to identify tumours in specific regions of a mammogram. We must remember that the dataset chosen contains images of full-breast mammograms. It could therefore be possible to use the Guided Backprop method in detailed sectioned mammograms, to identify the source of cancer.

Similarly, the LRP Preset A-Flat method was found not to be the best method to determine those neurons which contribute most to the classification of a mammogram. For this method, we tested out different betas and epsilon stabilisers, in an attempt to increase the clarity of the output image. The best result we got was when using a beta of 1 and an epsilon stabiliser of 0.07. Once this optimum was determined, all images within the subset which were processed using these values. The results obtained, although improved upon the Guided Backprop method were not sufficient for our purposes. The reason behind this conclusion is that the A-Flat method was able to show the outline of the breast within the mammogram, but failed to highlight those areas where cancer could be found. Nonetheless, there were instances in the subset where abnormalities were detected, pointing to the presence of major metastatic breast cancer; such as in cases 016, 026, 063, 172 and 242. In all instances the A-Flat method enlarged those pixels which showed an abnormality, so it would be beneficial as a first step to identify those mammograms which show a sign of abnormality, but these are also visible to the naked eye, thus not contributing that much to the analysis.

\begin{figure}[ht]
	\centering
	\includegraphics[width=1\textwidth]{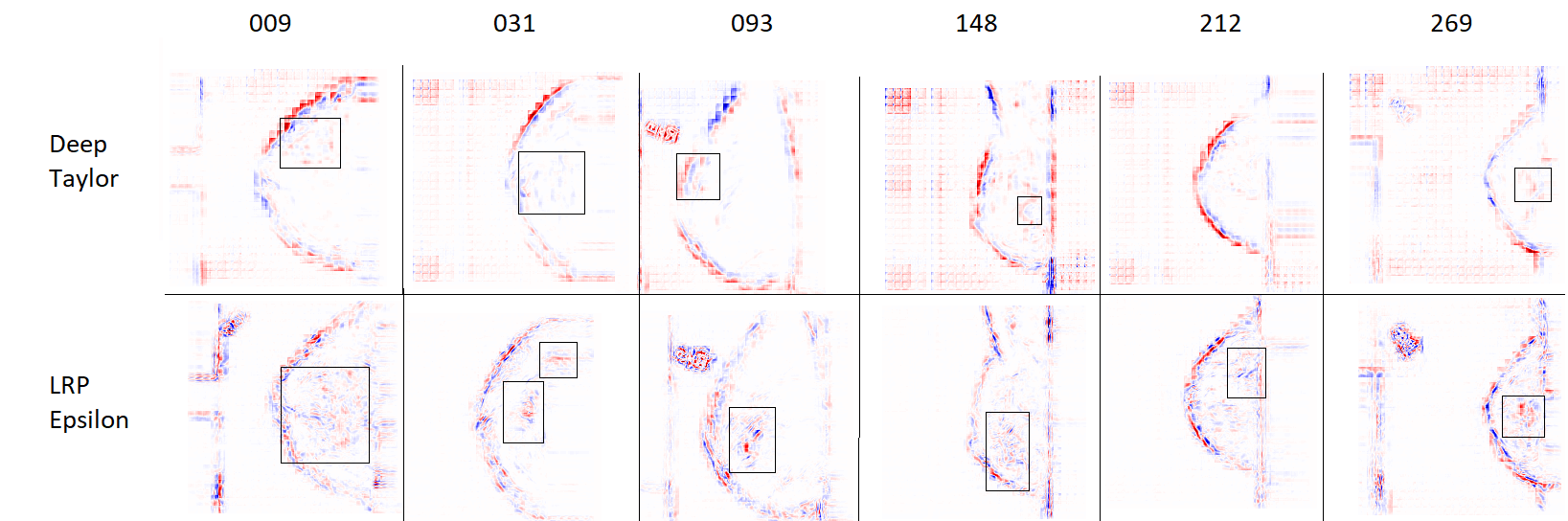}
	\caption[We gave specific importance to case numbers 009, 031, 093, 148, 212 and 269, which are shown in this order. In these mammograms, the Epsilon method showed clearer pixels than the Deep Taylor Decomposition method, however, that does not mean that they are always better in interpreting mammograms.]{We gave specific importance to case numbers 009, 031, 093, 148, 212 and 269, which are shown in this order. In these mammograms, the Epsilon method showed clearer pixels than the Deep Taylor Decomposition method, however, that does not mean that they are always better at interpreting mammograms.}        
\end{figure}

After having analysed and eliminated two of the five options using a visual comparison made using the naked eye, we remained with the LRP Epsilon, DeConvNet and the Deep Taylor Decomposition methods, to help us understand how the neural network classified the mammograms. All three methods provide relevant information about the input images, so our decision and evaluation had to be based on probability. The question we asked ourselves when evaluating the results of these three methods is which of them has the highest probability of identifying abnormalities in a mammogram. Based on the subset of 20 mammograms, we noticed that the Deep Taylor Decomposition method performed better than the other two in 12 of the 20 cases (016, 026, 030, 056, 063, 072, 111, 172, 178, 181, 223 and 259). In another 6 of the 20 images taken in the subset, it was noted that the Epsilon method generated results which visually showed more regions of interest than the ones generated by the Deep Taylor Decomposition. (009, 031, 093, 148, 212 and 269). These can be visualised in Figure 1. In one particular case; for case no: 131 it was also noted that the image was not clear enough to determine a classification. The results obtained additionally showed that the images being generated by the DeConvNet method were inferior to the ones generated by the Deep Taylor and Epsilon methods. These generated images contained more noise, thus losing pixel-wide information, when it comes to details.

\begin{figure}[ht]
	\centering
	\includegraphics[width=1\textwidth]{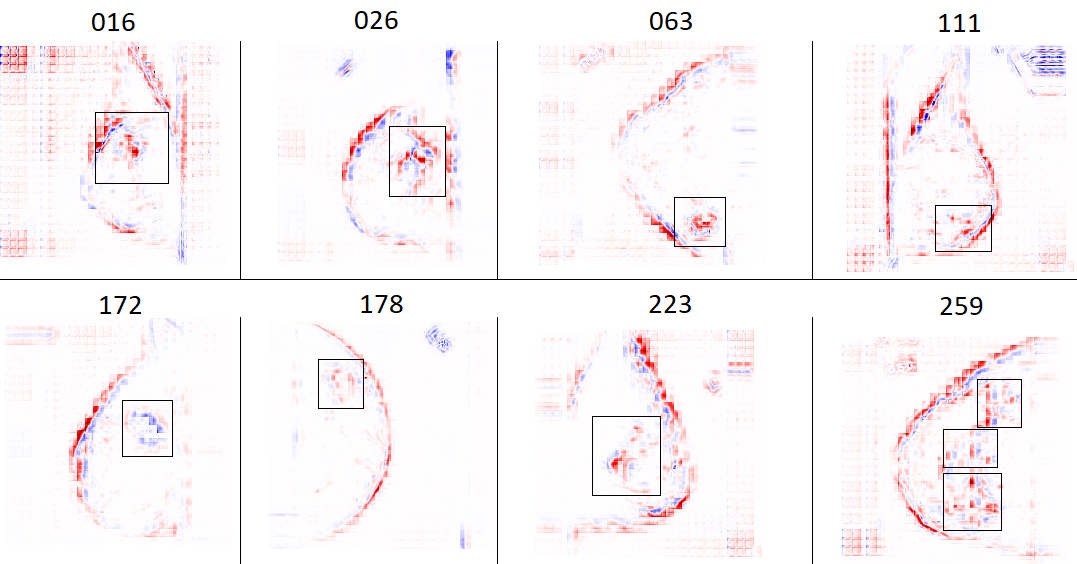}
	\caption[We used the iNNvestigate library Deep Taylor Decomposition method to identify those areas within a source image which have mostly contributed to the classification of a said mammogram. In the cases shown here, we are identifying those cases where the Deep Taylor Decomposition Method identified instances of malignant breast cancer.]{We used the iNNvestigate library Deep Taylor Decomposition method to identify those areas within a source image which have mostly contributed to the classification of a said mammogram. In the cases shown here, we are identifying those cases where the Deep Taylor Decomposition Method identified instances of malignant breast cancer.}        
\end{figure}

We focused for a moment on those cases where the Epsilon method generated results which showed more regions of interest to the naked eye than when using Deep Taylor Decomposition. Figure 1 shows the results obtained by each of these six cases in particular. The figure shows the results obtained by the Deep Taylor Decomposition method in the first column and the respective result obtained by the Epsilon method to the right. For case 009, The Epsilon results show a larger area which could contribute to the classification of the mammogram. The results obtained by the Deep Taylor Decomposition show a more specific detected area of concern. In this particular case, when comparing to the original mammogram, we noticed that the larger area detected by the Epsilon method could result from the presence of breast milk contained within the lobes of the patient's breast. So in this case, the Deep Taylor Decomposition method proved to be better in identifying the area of a concern than the Epsilon method. When comparing the two methods on case numbers 031 and 223, the Deep Taylor Decomposition found one single large area of concern, while the Epsilon method figured out two separate ones. For case number 093, the Epsilon method again performed better than the Deep Taylor Decomposition one, since it identified a better concerning area. In these two instances, we can see that the Epsilon method had superior performance to the Deep Taylor one, highlighting the affected areas in a better way. For cases 148 and 269, we noticed that while the pixels were clearer when using the Epsilon method, the images on the whole gave a better interpretation of the affected areas when using the Deep Taylor Decomposition method.

\begin{figure}[ht]
	\centering
	\includegraphics[width=0.8\textwidth]{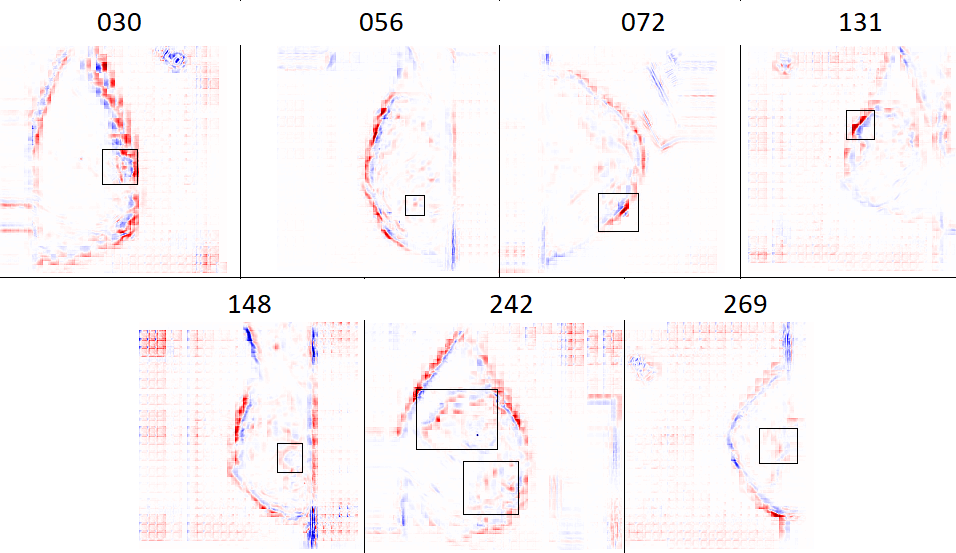}
	\caption[The Deep Taylor Decomposition method was able to also identify instances of benign breast cancer by highlighting those areas within a mammogram which contribute to its classification by using heatmaps.]{The Deep Taylor Decomposition method was able to also identify instances of benign breast cancer by highlighting those areas within a mammogram which contribute to its classification by using heatmaps.}        
\end{figure}

Based on the results obtained, we were able to understand which areas within a mammogram contribute to its classification into a breast containing, no signs of cancer, having benign cancer or having signs of malignant cancer. It became clear to us that cases no: 016, 026, 063, 111, 172, 178, 223 and 259 for example had a clear signs of malignant cancer. The Deep Taylor Decomposition method was able to successfully show the affected areas of the breast by highlighting them in either blue or red colours. These can be seen in Figure 2. The same method was also able to define those mammograms which may have a presence of benign breast cancer in case numbers 030, 056, 072, 131, 148, 242 and 269. These can be seen in Figure 3. This is where the assistance of a radiologist would be needed to help in the diagnosis of such cases. It was also noted that in cases 009, 031, 093 and 181 the Deep Taylor Decomposition method was not able to find any signs of breast cancer; as shown in Figure 4. Questionable among these is case no: 093, whereby although the Deep Taylor method did not find any abnormalities, the Epsilon method did. This is shown in Figure 5.

\begin{figure}[ht]
	\centering
	\includegraphics[width=0.8\textwidth]{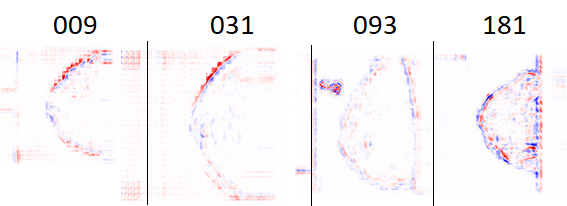}
	\caption[In some cases, the Deep Taylor Decomposition method we used was not able to identify the presence of breast cancer. While the model took this decision, we still concluded that these mammograms should be interpreted also by a human radiologist.]{In some cases, the Deep Taylor Decomposition method we used was not able to identify the presence of breast cancer. While the model took this decision, we still concluded that these mammograms should be interpreted also by a human radiologist.}        
\end{figure}

\begin{figure}[ht]
	\centering
	\includegraphics[width=0.5\textwidth]{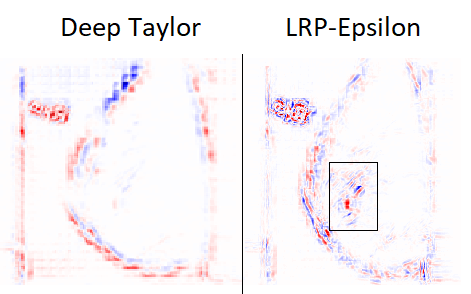}
	\caption[In one particular case, where the Deep Taylor Decomposition method was unable to detect any areas pertaining to a potential classification of benign or malignant cancer, The Epsilon method was able to identify an area which potentially could contain such signs of cancer.]{In one particular case, where the Deep Taylor Decomposition method was unable to detect any areas pertaining to a potential classification of benign or malignant cancer, The Epsilon method was able to identify an area which potentially could contain such signs of cancer.}        
\end{figure}

\section{Evaluation}

To assist in the evaluation of the developed XAI model, we created a second questionnaire and shared it with the same 12 clinical specialists from Malta who were chosen for the initial questionnaire mentioned in the introduction. This time, the response received involved less participants, but the study was more focused and detailed such that we decided to meet three specialists in breast radiology, and two in breast surgery individually to get a first-hand evaluation from them on the proposed approach and model. 

The initial comments we received from the interviewed specialists was that the chosen dataset could have been a better one. Even though the \textit{CBIS-DDSM} dataset is a widely used public dataset of breast mammograms which have been thoroughly curated and annotated by experts in the field, it was noted that the quality of the images was inferior to the ones they were accustomed to. The amount of detail which the \textit{BI-RADS} mammograms go into can therefore be much more than the ones processed by \textit{DICOM} machines, therefore allowing for microcalcifications in the breast to be identified by radiologists in an easier way.

The second point which the clinicians noted was the quality of the mammograms in the \textit{CBIS-DDSM} dataset. In particular case numbers: 072, 056, 063, 148 and 259 were ruled as being unidentifiable, and stating that if they were presented with such mammograms, they would ask for the radiographer to perform a second mammographic test on the patient.

A third point which the clinicians mentioned in the interview was that the \textit{CBIS-DDSM} dataset only displays a one-sided full-size mammogram of each case. Tomosynthethic images have a three dimensional model of each breast, by taking multiple mammographic images of the same breast. Therefore what could be seen as a microcalcification from one facet of the mammogram, could be further viewed from different facets of the same breasts and diagnosed in a better way \cite{Ekpo}.

\section{Conclusions}

From the results obtained in the preliminary study made during the summer months of 2020. We can conclude that even though the contribution and involvement of deep learning algorithms were generalised and toned down, the majority of participants who participated in the questionnaire increased their element of trust when the idea of XAI was introduced. The imposition of a back propagation model which highlights those pixels in a deep learning model contributes to its classification, and visualising these areas through a heatmap would indeed provide a better solution to the conventional CNN models. Knowing how a model has classified different mammograms into containing benign, malignant or no signs of a tumour would be beneficial to radiologists who are attempting to diagnose the presence of breast cancer in a patient. 

Having said that, it remains a fact that the specialist showed concern around the use of the \textit{CBIS-DDSM} dataset for this study and its lack of detail when compared to the mammograms which the specialists are accustomed to at Mater Dei Hospital. The interviewed participants also commented about the four elements which radiologists look into when studying a mammographic image; being the presence of microcalcifications, an asymmetry between the two breasts, distortion in particular areas and opaqueness of that area.

The combination of a ResNet-50 architectured model together with the Deep Taylor Decomposition methodology has theoretically proved to be a good combination to build a deep learning model and apply backpropagation to it. However, in the scenario set by this study and as expressed by the medical specialists, improvements need to be made to the dataset for a better evaluation of the XAI model. The suggestion given by the radiologists interviewed in the second part of the study was to choose particular mammograms from the \textit{CBIS-DDSM} dataset and run the XAI model against them. This will help evaluate the effectiveness of the XAI model in a better way.  

As Ayhan et al. discuss in their paper about guided backdrop methods and their performance in relation to the Deep Taylor decomposition methods, the former seems to generate saliency maps which are better in explaining decisions taken by Deep Learning models. The study concluded that the Guided Backprop method is still in its initial stages when attempting to read from a ResNet-50 model. Therefore classification predictions on medical images cannot be considered conclusive. The reason for this according to Ayhan et. al is due to the restrictions which the Guided Backprop model by design has with ReLU activators. The model should be extended to new architectures beyond the ReLU activator-based models \cite{ayhan2021clinical}. Nonetheless, the Guided Backprop approach was never employed as a saliency map to reverse engineer mammograms used in breast cancer interpretation. So it remains to be seen whether Guided Backprop can uncover those mammography pixels that contribute the most to the ultimate assessment of benign, cancerous, or normal breast tissue in a future investigation. It is also necessary to demonstrate if this strategy works well with the ResNet-50 patch-classifier models or whether an alternative model should be used to develop a CNN that can successfully diagnose breast cancer.

\bibliographystyle{splncs04}

\bibliography{paper4}

\end{document}